\documentstyle[aps,epsfig]{revtex}
  
\begin{document}  
\draft 
\title{
Left-right asymmetries, the weak mixing angle, and new physics}
\author{J. C. Montero, V. Pleitez and M. C. Rodriguez} 
\address{
Instituto de F\'\i sica  Te\'orica\\ 
Universidade  Estadual Paulista\\
Rua Pamplona, 145\\ 
01405-900-- S\~ao Paulo, SP\\ 
Brazil.} 
\maketitle 
\begin{abstract}  
The goal of this article is to outline the advantages of the measurement of 
left-right asymmetries in lepton-lepton ($l^{-}l^{-}\to l^{-}l^{-}$) scattering
for performing precision measurements of $\sin^2 \theta_W$ and the discovery of ``new
physics''. 

\end{abstract}
\pacs{PACS   numbers: 13.88.+e; 
12.60.-i 
12.60.Cn;  
}
   
\section{Introduction}
\label{sec:intro}

It is well known that the observables in the lepton-lepton scattering have less
uncertainties than in the lepton-hadron or hadron-hadron cases. 
This is because gauge models only specify the lepton-quark or quark-quark
vertices, and some parton model assumptions of hadron structure must be invoked
to relate the lepton-quark and quark-quark interactions with the lepton-hadron
and hadron-hadron ones, and this implies in introducing some uncertainties in
the calculation. 

In particular, the appealing features for studying the parity-violating
asymmetries between the scattering of left- and right-handed polarized electrons
on a variety of fixed targets ($e^-e^-, e^-\mu^-$) were pointed out some years
ago by Derman and Marciano~\cite{dm}, and they were systematically studied 
in lepton-lepton scattering for both fixed target ($e^-e^-, e^-\mu^-$) and
collider ($e^-e^-, e^-\mu^-, \mu^-\mu^-$) experiments in
Refs.~\cite{pg,cm1,assi1,assi2,mex1}.   

The left-right asymmetry is defined as
\begin{equation}
A_{RL}(ll\to ll)=\frac{d\sigma_R-d\sigma_L}{d\sigma_R+d\sigma_L},
\label{a1}
\end{equation}
where $d\sigma_{R(L)}$ is the differential cross section for one right
(left)-handed lepton $l$ scattering on an unpolarized lepton $l$.
Another interesting possibility is the case when both leptons are 
polarized. We can define an asymmetry $A_{R;RL}$ in which
one beam is always in the same polarization state, say right-handed, and 
the other  is either
right- or left-handed polarized (similarly we can define  $A_{L;LR}$):
 \begin{equation}
A_{R;RL}=\frac{d\sigma_{RR}-d\sigma_{RL}}{d\sigma_{RR}+d\sigma_{RL}},\qquad
A_{L;RL}=\frac{d\sigma_{LR}-d\sigma_{LL}}{d\sigma_{LL}+d\sigma_{LR}}.\qquad
\label{a2}
\end{equation}

We can define also an asymmetry when one incident particle is right-
handed and the other is left-handed and the final states are right- and left or
left- and right-handed:
\begin{equation}
A_{RL;RL,LR}=\frac{d\sigma_{RL;RL}-d\sigma_{RL;LR}}{d\sigma_{RL;RL} 
+ d\sigma_{RL;LR}},
\label{a3}
\end{equation}
or similarly, $A_{LR;RL,LR}$. All of these asymmetries can be calculated for
both fixed target and colliders  experiments. For more details on the
notation see Ref.~\cite{assi1}.

The appealing features of this type of measurement, following~\cite{dm}, are:
${\bf (1)}$ The asymmetry is manifestly parity violating: therefore its
measurement determines the electron's parity violation weak neutral current
interaction with the target or with the other beam in the case of colliders.
${\bf (2)}$ Because the effect investigated is due to the interference between
the weak and electromagnetic amplitudes, the asymmetry is proportional to $G_{F}$
and hence larger than the usual weak interaction effects which are ${\cal
O}(G^{2}_{F})$. ${\bf (3)}$ Since the asymmetry is a ratio, uncertainties
(theoretical and experimental) which are common to both the numerator and
denominator cancel out and therefore we can use them to perform precision
measurements in the context of the standard model~\cite{gws} (see
section~\ref{sec:precision}). 
${\bf (4)}$ Finally, this kind of interference measurements determines the
relative sign between the weak and electromagnetic interactions. Unified 
gauge theories give unique predictions for this algebraic sign; therefore their 
determination provides an additional check on various models- i.e. some models 
may predict the correct magnitude for the asymmetries, but the wrong sign!
(see section~\ref{sec:new}). 
For fixed target experiments, in $e^-e^-$ scattering a very intense polarized 
electron beam inciding on an unpolarized hydrogen target while in $e^-\mu^-$
scattering it is considered an inciding unpolarized muon beam on polarized
electron target.  

There are some advantages in considering asymmetries in $e^-e^-$ and $e^-\mu^-$
scattering. Firstly, although in fixed target experiment the value of $A_{RL}$
asymmetry is small (see below), the cross sections of these processes are large
allowing for a good statistic and, on the other hand, in collider experiments
the asymmetry is large but cross sections are small.
Secondly, in $e^{-} \mu^{-}$ scattering the background 
contribution from $\mu^{-}N$ scattering is less severe because one could 
trigger on a single scattered electron which could not have arisen from a 
$\mu^{-}N$ collision. Besides in the $e^-\mu^-$ case that muon beams generally
have energies at least one order of magnitude greater than the electron beams. 
We would like to point out that these asymmetries could be already measured in
existing fixed target experiments like E158~\cite{e158} and NA47~\cite{smc}.

On the other hand, we have being interested in these asymmetries in collider
experiments since in the future we hope that colliders like the so called Next Linear Collider
(NLC)~\cite{nlc} or the International Linear Collider (ILC)~\cite{exp1} could
work in the $e^-e^-$ mode and the First Muon Collider (FMC)~\cite{gunion}
could work in $\mu^{-} \mu^{-}$ modes and hybrid collider could do well
in $e^{-} \mu^{-}$ scattering.

This paper is organized as follows. The usefulness of the measurements of these
asymmetries in collider and fixed target experiments are discussed briefly in
section~\ref{sec:precision}; while ``new physics'' in section~\ref{sec:new}. Our
conclusion are given in the last section, \ref{sec:con}.

\section{Determination of $\sin^2 \theta_W$}
\label{sec:precision}

At present only a couple of observables differ from the 
prediction of the standard model at the level of 3 standard
deviations~\cite{ewwg}: the value of the forward-backward asymmetry
$A^{0,b}_{{\rm fb}}$ in $e^+e^-$, and the value of $\sin^2\theta_W({\nu N})$
obtained from the ratios of the charged and neutral currents in neutrino-nucleon
scattering~\cite{nutev}. Hence a better measurement of the weak mixing angle in
process different from $e^+e^-$ and $\nu N$ scattering has become
mandatory. This could be the left-right asymmetries in $e^-e^-$. 

In the standard electroweak model the $A_{RL}(e^{-}e^{-})$ asymmetry at the
tree level in fixed target experiment is given by:
\begin{equation}
A^{{\rm FT;SM}}_{RL}(ee)\approx-\frac{G_{F} Q^{2}}{\sqrt2\pi\alpha}\,
\frac{1-y}{1+y^{4}+(1-y)^{4}}\,(1-4 \sin^{2} \theta_{W}),
\label{assi0}
\end{equation}
where $Q^{2}=-y(2m^{2}_{e}+2m_{e}E_{\rm beam})$, $y= \sin(\theta/2)$.
In the Eq.~(\ref{assi0}) the approximation $m^{2}_{e} \ll Q^{2}\ll M^{2}_{Z}$
was used, and if $E_{\rm beam}=50$ GeV, $\theta\approx90^o$ ($y=1/2$) we obtain the
following value~\cite{dm,assi1}
\begin{equation}
A^{{\rm FT;SM}}_{RL}(ee)\approx-3\times 10^{-7}.
\label{aft1}
\end{equation}
Marciano and Czarnecki, Ref.~\cite{cm} 
have calculated the one loop electroweak radiative corrections and found a 
rather substantial $40 \pm 3 \%$ reduction of the tree level prediction.

We have also verified that $A_{RL}(e^{-} \mu^{-})$, in fixed target experiments, 
is sensitive to the value of $\sin^2 \theta_W$. In this case the $A_{RL}$
asymmetry, for $m_{\mu}=0$, is given by~\cite{dm,assi2}:
\begin{equation}
\left.A_{RL}^{FT;SM}(\mu e)\right\vert_{m_{\mu}=0}\approx
-\frac{8G_{F}}{\sqrt{2}}\; 
\frac{g_{V}g_{A}\,M^{2}_{W}}{ \pi \alpha M^{2}_{Z}}\;  
\frac{ys}{1+(1-y)^{2}}\;(1-4 \sin^{2}\theta_{W}).
\label{asydm}
\end{equation}
If we consider a fixed target experiment with $E_{\mu}=190$ GeV, $E_{\mu}$ is 
the muon beam energy, a 0.5\% change in the $A_{RL}(e^{-} \mu^{-})$  value 
corresponds to a 0.04\% change in  $\sin^{2}(\theta_{W})$, {\it i.e.}, a change 
from $0.2315$ to $0.2316$~\cite{assi2}. This fact show that it could be useful
for doing very precise electroweak studies in this kind of experiment too. 

\begin{figure}[ht]
\begin{center}
\vglue -0.009cm
\mbox{\epsfig{file=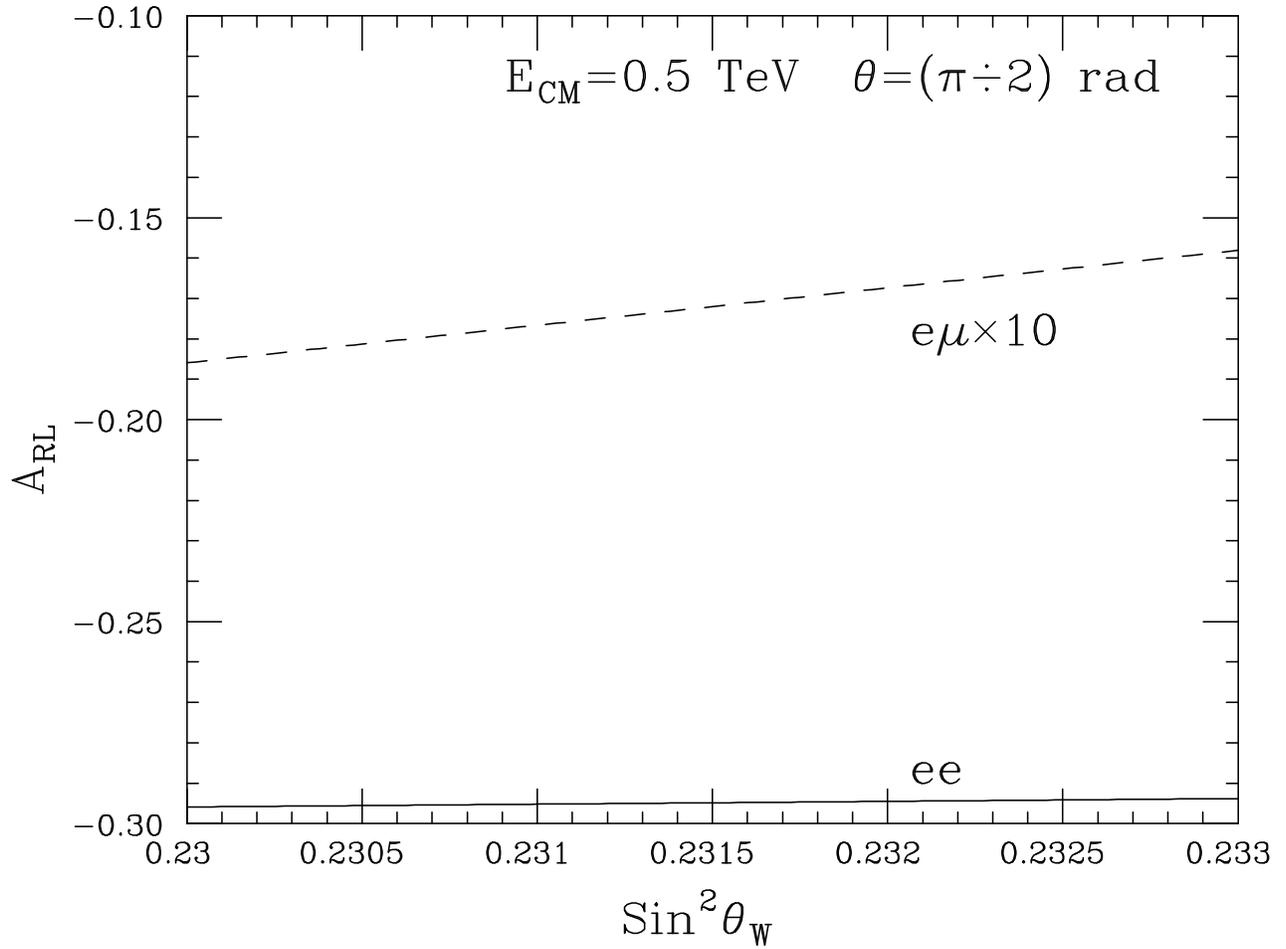,width=0.7\textwidth,angle=90}}       
\end{center}
\caption{The $A_{RL}$ asymmetry as a function of $\sin^2\theta_W$ for $e^-e^-$
and $e^- \mu^-$ collider experiments and $E_{\rm CM}=\sqrt{s}=0.5$TeV.}
\label{f1}
\end{figure}

More recently it was pointed out in Ref.~\cite{cm1} that the asymmetry
$A_{RL}(e^{-}e^{-})$ in colliders experiments can be used to measure $\sin^{2}
\theta_{W}$ rather precisely. Here we will briefly discuss the
$A_{RL}$ asymmetry for colliders experiments, for more details see
Refs.~\cite{assi1,assi2}.   
In Fig.~\ref{f1} we plot the $A_{RL}$ as function of $\sin^{2} \theta_{W}$ to
both $e^{-}e^{-}$ and $e^{-} \mu^{-}$ colliders. 
The results for other values of $E_{\rm CM}$ are shown in Table~\ref{t1}. 
We see that the weak mixing angle is more sensitive to that asymmetry for
the  $e^-\mu^-$ scattering.  

It is clear that in the case of
$e^{-}e^{-}$ we need to measure $A_{RL}$ with more precision than in the case 
$e^{-} \mu^{-}$ to have the same variation in $\sin^{2} \theta_{W}$ at the 
energies $\sqrt{s}=1.0 \,\ {\mbox TeV},1.5 \,\ {\mbox TeV}$ and $2$ TeV. In the 
case of $\sqrt{s}=0.5$ TeV we need the same precision in both experiments. We 
should stress that the results for $\mu^{-} \mu^{-}$ scattering are the same as
in the case of  $e^{-}e^{-}$ \cite{assi3}. 

As we have shown on this section the standard model implies a predictable degree
of parity violation in lepton-lepton scattering, ranging from low energy
phenomena to high energy, and we can use them to do very precise tests of the 
model. Of course, such sensitivity implies that a measurement of 
$A_{RL}$ is also a good probe of ``new physics'' if it really exists, as we 
will show on the next section. 

\section{New Physics}
\label{sec:new}

In general the standard model is exceedingly successful in describing leptons, quarks 
and their interactions, it is in excellent agreement with the worldwide 
data~\cite{pdg}. However, the necessity to go beyond it, from the 
experimental point of view, comes at the moment only from neutrino data 
\cite{neu}. If neutrinos are massive then new physics beyond the standard 
model is needed.

On the other hand, any extension of the standard model implies 
necessarily the existence of new particles. We can have a rich scalar-boson 
sector if there are several Higgs-boson multiplets~\cite{higgshunter} or have 
more vector and scalar fields in models with a larger gauge symmetry as in 
the left--right symmetric~\cite{lr} and in 3-3-1 models~\cite{331}, or we 
also  can have at the same time more scalar, fermion, and vector particles 
as in the supersymmetric extensions of the standard model~\cite{mssm}, or in the
supersymmetric 3-3-1 model~\cite{susy331}. 

Recently Dimopoulos and Kaplan~\cite{dk} take up again the Weinberg's
idea~\cite{sw72} that the observed value of the weak mixing angle
$\sin^2\theta_W=.231$ suggests indeed an $SU(3)$ electroweak symmetry, for
instance models with $SU(3)\otimes SU(2)\otimes U(1)_N$ electroweak gauge
symmetry at the 1 TeV scale. The point is that in such models
$\sin^2\theta_W(M)=1/4$, with $M$ being an energy scale of $SU(3)$ breaking.
This is very important since it means the possible existence of a new
fundamental energy scale which is not related to supersymmetry,
neutrino masses, unification, or superstrings~\cite{mdine}. At this energy scale
new exotic quarks, doubly charged vector bosons and/or extra dimensions may
exist. One way of considering the known quarks with the extra $SU(3)$
electroweak symmetry model is by embedding it in a Pati-Salam-type
model~\cite{ps}, with gauge symmetry $SU(4)_c\otimes SU(2)^\prime_L\otimes 
SU(2)^\prime_R\otimes SU(3)^\prime$, and with the known quarks transforming only
through the first three factors~\cite{dk}. 
There is a first symmetry breakdown to $SU(3)_c\otimes SU(2)^\prime_L\otimes
U(1)^\prime$, and finally to $SU(2)_L\otimes  U(1)_Y$. Another interesting way
to solve the introduction of quarks without invoking a $U(1)$ factor as in the
3-3-1 models~\cite{331}, is to
consider $SU(3)_c\otimes SU(3)_W$ models in 5 dimensions~\cite{335d}.

Independently of this interesting fact, we recall that the $SU(3)$ symmetry
among the lightest particles i.e., $e^-$, $e^+$ and $\nu_e$ could be the last
symmetry involving the known leptons or $SU(4)$ if we add right-handed
neutrinos~\cite{su4,foot}. If we impose this symmetry on the electron sector 
we must also to impose it upon all the other particles and if we do not want to
introduce extra dimensions we must introduce extra exotic charged quarks. In
this case it is also possible to have anomaly cancellation only if the number of
families is three or a multiple of three~\cite{331}. If we accept that the
observed value for the weak mixing angle is an indication of a $SU(3)$ symmetry
at the TeV level, it means that the respective supersymmetric model has also to
have this $SU(3)$ symmetry, and supersymmetry has to be naturally broken at the
same scale and it can also be embedded in theories of TeV-gravity~\cite{dk}.
Any way, all those sort of models involve doubly charged vector
bosons. 

Usually we have several particles contributing to a given observable,
and for this reason it is very difficult to identify their contributions in
both the usual and exotic processes. 
In some models~\cite{331,susy331,derli} the
contributions of the scalar-bosons are not suppressed by the fermion masses 
and they can have the same
strength of the fermion--vector-boson coupling. However, for $e^-e^-$ and
$\mu^-\mu^-$ scattering, in the $s$-channel, the contributions to the
$A_{RL}$ asymmetry of doubly charged scalars, like $H^{--}$, cancel out. 
Hence these parity violating asymmetries are only sensitive to the doubly
charged vector bosons. 
In view of this, we will discuss in this section the role of $A_{RL}$
asymmetry in finding signal for ``new physics'': doubly charged and neutral 
vector bosons, $U^{--}_\mu$ and $Z^\prime$, respectively~\cite{mex2}. 

In Refs.~\cite{assi1,assi3} it was noted that the left-right asymmetries in the
lepton--lepton diagonal scattering are quite sensible to doubly-charged vector
boson $U$ contributions.  The main result obtained is 
\begin{center}
\begin{equation}
A^{\rm CO;ESM+U}_{R;RL}(l^{-}l^{-} \to l^{-}l^{-}) \approx - 
A^{\rm CO;ESM}_{R;RL}(l^{-}l^{-} \to l^{-}l^{-}),
\end{equation}
\end{center}
where $l=e, \mu$.

In the model of the Ref.~\cite{331} there is also a $Z^{ \prime}$ neutral vector boson 
which couples with leptons. It was shown in Ref.~\cite{assi2} that for the
non-diagonal scattering ($e^{-} \mu^{-}$) the asymmetries are sensible to the 
existence of an extra neutral vector-boson $Z^{ \prime}$, because
\begin{center}
\begin{equation}
A^{\rm CO;ESM}_{RL}(e^{-} \mu^{-} \to e^{-} \mu^{-}) \neq A^{\rm
CO;ESM+Z^{ \prime}}_{RL}(e^{-} \mu^{-} \to e^{-} \mu^{-}).
\end{equation}
\end{center}

Hence, both $U^{++}$ and $Z^{ \prime}$ vector bosons can be potentially
discovered in these sort of processes by measuring the left-right asymmetries. 
Here we should stress that although the results about the $U^{++}$ and 
$Z^{ \prime}$ were obtained in the context of the 3-3-1 model they are still
valid in the supersymmetric version of the latter model~\cite{susy331}. The
reason is that the lagrangian of the interaction between leptons and gauge
bosons are the same in both models~\cite{cr}.

Finally, we want to stress that the $e^{-}e^{-}$ collider is 
ideal for discovering and studying selectrons. This will allow to do precise
measurements of the neutralino masses and the opportunity to observe and analyze
cascade decays of the selectron~\cite{cu1,cu2}.

\section{Conclusions}
\label{sec:con}

We should remember that left-right asymmetries have played key roles in 
establishing the validity of the standard model. They will continue to provide 
valuable tools during the next generation of colliders doing precision studies 
of the standard model and even more exciting is the possible direct detection 
of new phenomena such as bileptons, $Z^{\prime}$s, supersymmetry, and others.
Here in this work we have reviewed how the asymmetries, defined in
Ref.~\cite{assi1} can be used to perform this kind of study.

\acknowledgments 
This work was supported by Funda\c{c}\~ao de Amparo \`a Pesquisa
do Estado de S\~ao Paulo (FAPESP), Conselho Nacional de 
Ci\^encia e Tecnologia (CNPq) and by Programa de Apoio a
N\'ucleos de Excel\^encia (PRONEX).


\begin{table}
\caption{Dependence of the $A_{RL}$ asymmetry on the $\sin^2\theta_W$. }
\begin{tabular}{||c rr||}
\hline
$\sin^{2} \theta_{W}$ & $A^{CO;ESM}_{RL}(e^-e^-)$  &
$A^{CO;ESM}_{RL}(e^-\mu^-)$\\ 
\hline
& $E_{CM}=0.5$ TeV & \\
0.23073 & -0.29536 & -0.01789 \\
0.23063& -0.29545 & -0.01801 \\ 
\hline
& $E_{CM}=1.0$ TeV & \\
0.23073 & -0.33309 & -0.01977 \\
0.23063 & -0.33319 & -0.01989\\
\hline
& $E_{CM}=1.5$ TeV & \\
0.23073 & -0.34786 & -0.02016 \\
0.23063 & -0.34796 & -0.02028 \\
\hline
& $E_{CM}=2$ TeV & \\
0.23073 & -0.35574 & -0.02030 \\
0.23063 & -0.35583 & -0.02043
 \end{tabular}
\label{t1}
\end{table}

\end{document}